\documentclass[aps,prl,twocolumn,superscriptaddress,showpacs,floatfix]{revtex4}
\usepackage{graphicx}
\usepackage{hyperref}
\usepackage{amsmath}
\usepackage{color}

\newcommand{\ket}[1]{|{#1}\rangle} 

\newcommand{\ud}[1]{{#1^{\dagger}}}

\begin{document}
\flushbottom
\title{Exciting polaritons with quantum light}

\author{J.~C.~L\'opez Carre\~no}
\affiliation{Departamento de F\'isica Te\'orica de la Materia  Condensada and Condensed Matter Physics Center (IFIMAC), Universidad Aut\'onoma de Madrid, E-28049, Spain.}
\author{C.~S\'anchez Mu\~noz}
\affiliation{Departamento de F\'isica Te\'orica de la Materia Condensada and Condensed Matter Physics Center (IFIMAC), Universidad Aut\'onoma de Madrid, E-28049, Spain.}
\author{D.~Sanvitto}
\affiliation{Nanotec, Istituto di Nanotecnologia-CNR, Via Arnesano, 73100 Lecce, Italy}
\author{E.~del Valle}
\affiliation{Departamento de F\'isica Te\'orica de la Materia Condensada and Condensed Matter Physics Center (IFIMAC), Universidad Aut\'onoma de Madrid, E-28049, Spain.}
\author{F.~P.~Laussy}
\affiliation{Russian Quantum Center, Novaya 100, 143025 Skolkovo, Moscow Region, Russia}
\affiliation{Departamento de F\'isica Te\'orica de la Materia Condensada and Condensed Matter Physics Center (IFIMAC), Universidad Aut\'onoma de Madrid, E-28049, Spain.}
\email{fabrice.laussy@gmail.com}

\begin{abstract}
  We discuss the excitation of polaritons---strongly-coupled states of
  light and matter---by quantum light, instead of the usual laser or
  thermal excitation. As one illustration of the new horizons thus
  opened, we introduce ``\emph{Mollow spectroscopy}'', a theoretical
  concept for a spectroscopic technique that consists in scanning the
  output of resonance fluorescence onto an optical target, from which
  weak nonlinearities can be read with high precision even in strongly
  dissipative environments.
\end{abstract}
 \date{\today} \maketitle

\begin{figure}[t]
  \centering
  \includegraphics[width=.7\linewidth]{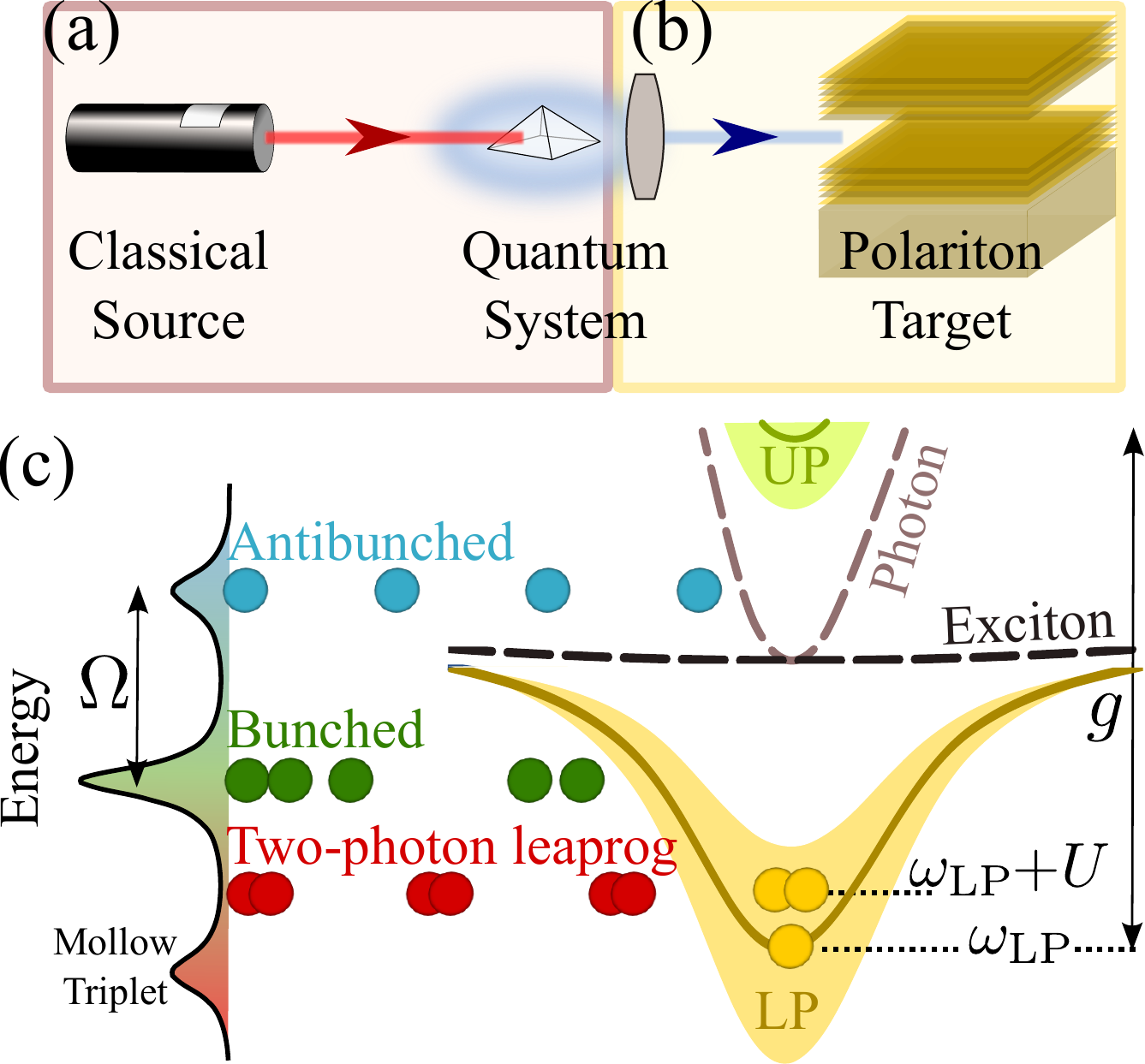}
  \caption{Scheme of our proposal: (a) a typical optical excitation
    scenario of quantum optics, a laser excites a quantum system,
    e.g., a quantum dot. (b) Instead of the conventional scenario of
    also exciting polaritons with a laser, we excite polaritons with
    quantum light, specifically, from the output of the quantum system
    excited by the laser. (c) Mollow spectroscopy: the
    photoluminescence of a strongly-driven two-level sytem provides
    the Mollow triplet, shown on the left with energy on the vertical
    axis. Various spectral windows provide different types of photon
    correlations, sketched here as photon balls with different
    temporal spacing. Exciting the lower polariton with leapfrog
    photon pairs allows to measure accurately very small values of the
    interaction.}
  \label{fig:1}
\end{figure}

Introduction. --- One of the chief concerns of modern optics is to
bring to fruition nontrivial quantum states of
light~\cite{obrien09a,bertolotti15a}. This is typically achieved by
driving a quantum system with a laser, turning light that is as
classical as can be according to quantum mechanics into a
non-classical output~\cite{dodonov02a,muller15a}.  In this Letter, we
take the reversed stand of driving a classical system with a quantum
source. By ``classical system'' is meant here one which would yield
classical states if excited by a laser, such as an harmonic oscillator
with Hamiltonian $H_a=\hbar \omega_a\ud{a}a$. This simple case is
still an important target as it describes, among other physical
systems of interest, the single mode of a passive cavity or a field of
non-interacting bosons such as
plasmons~\cite{javiergarciadeabajo12a}. To describe composite
particles such as exciton-polaritons~\cite{kavokin_book11a}, one then
simply considers two harmonic oscillators~$a$ and~$b$ linearly coupled
with strength~$g$. Excitons being weakly interacting, the polariton
Hamiltonian becomes anharmonic~\cite{verger06a}:
\begin{equation}
  \label{eq:lunabr27150838CEST2015}
  H_1=\omega_a\ud{a}a+\omega_b\ud{b}b+ g(\ud{a}b+a\ud{b})+ U\ud{b}\ud{b}bb\,.
\end{equation}
These systems are intrinsically open by nature and driving such
systems with a laser only allows for weak incursion~\cite{boulier14a},
if at all, into the quantum regime. This is due to interactions being
too small and dissipation too large for a laser to imprint genuinely
quantum features into the system and allow significant single-particle
effects to occur.  On the contrary, when driven by a quantum source,
even a linear system is left in a strongly quantum state.  This
motivates us in introducing the paradigm of exciting polaritons with
quantum light. This opens a new chapter of the field, already rich
with mesoscopic quantum states (such as
condensates~\cite{kasprzak06a}, superfluids~\cite{amo09b}, Josephson
oscillators~\cite{lagoudakis10a,abbarchi13a}, etc.)  that can then be
brought to the single-particle limit, with prospects of investigating
quantum simulators~\cite{byrnes11a,byrnes13a} or logic with polariton
Fock states~\cite{ballarini13a,anton13a}. We focus on two particular
applications to illustrate our approach and leave further applications
to future
works.

Formalism. --- With the ever increasing availability of quantum
emitters, the question of their effect on a target is one of
increasing theoretical interest. In the wake of the proposal for
quantum optical spectroscopy~\cite{kira06a}, it has been shown that
the statistics of light affects strongly the response of a
system~\cite{kira11a,carmele09a,assmann11b}. 
In our case, the excitation is the continuous wave (steady-state)
dynamical output from a quantum emitter, with its own theoretical
model and equations of motion.  The formalism would then appear to
simply demand to include the source as a part of the system and solve
for the dynamics of the joint exciting-excited components. An
important requisite, however, is that the source is unaffected by the
target: when an experimentalist shines light on a sample, the source's
internal dynamics remains in principle unaffected by the presence, or
not, of the sample. This is automatically realized in the conventional
model for excitation by a laser, since the latter is described by a
$c$-number, which has no internal dynamics. For instance, exciting a
two-level system~$\sigma$ with a laser is simply modeled by the
Hamiltonian~\cite{mollow69a}:
\begin{equation}
  \label{eq:lunabr20123138CEST2015}
  H_2=\omega_\sigma\ud{\sigma}\sigma+\Omega\exp(i\omega_\mathrm{L}t)\sigma+
  \Omega^{*}\exp(-i\omega_\mathrm{L}t)\ud{\sigma}
\end{equation}
and regardless of the population, coherence, etc., effectively
generated in the target, the attributes of the exciting laser
(intensity $|\Omega|^2$ and frequency~$\omega_\mathrm{L}$) remain fix
as they are mere parameters of the model. In the fully quantized
version~\cite{delvalle10d}, however, where the light field $\Omega
e^{i\omega_\mathrm{L}t}$ is upgraded to a Bose annihilation
operator~$a$ for the light field, there is now a feedback from the
emitter~$\sigma$ to the source~$a$. This causes some dynamics of the
supposed exciting laser, that affects, in turn, the target, removing
the fundamental asymmetry with the source that describes most
experimental setups.

\begin{figure}[t]
  \centering
  \includegraphics[width=\linewidth]{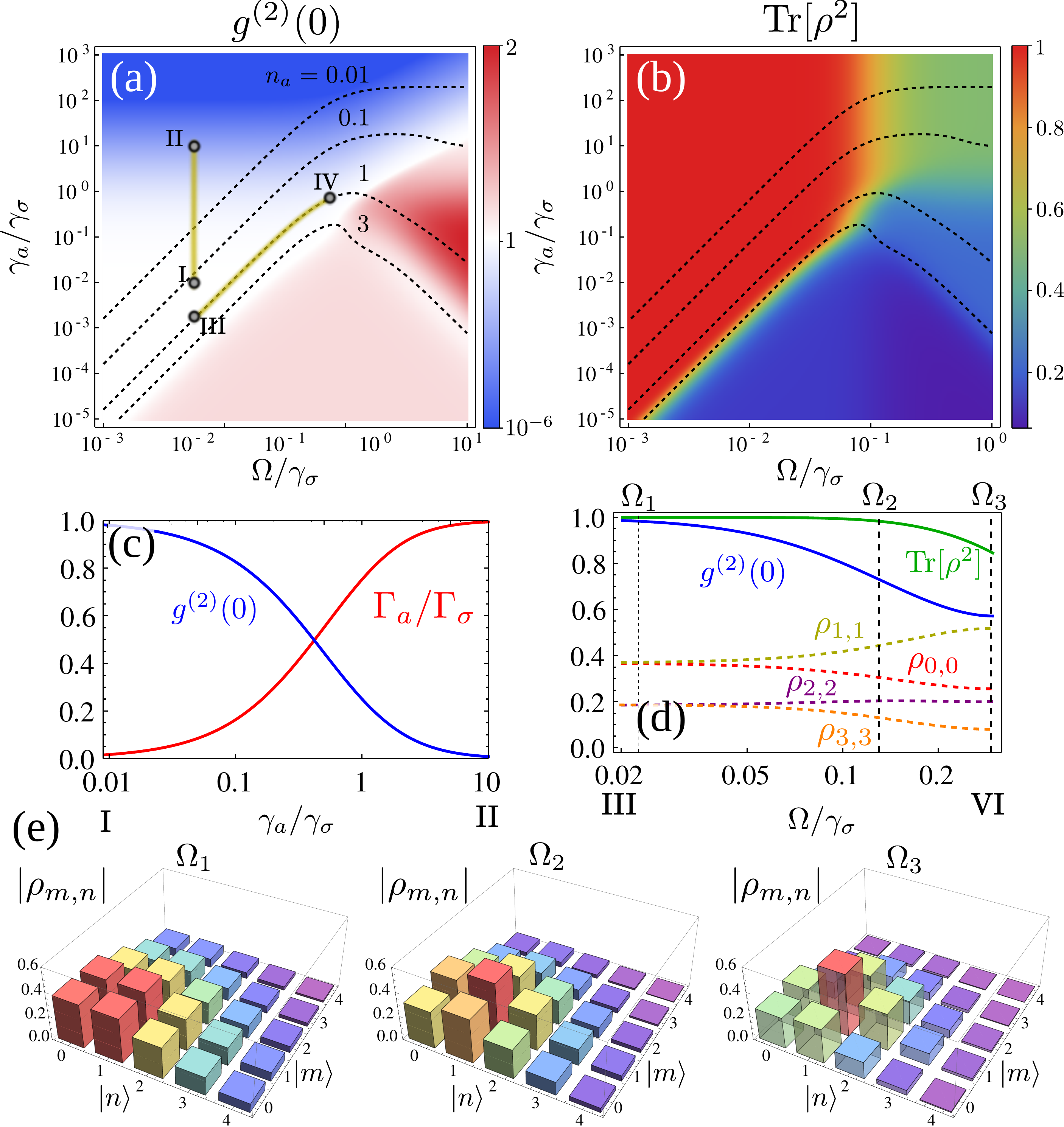}
  \caption{(Colour online) Pure quantum states in the steady
    state. The red area of (b) shows areas where the system is in a pure state. The density plot of~$g^{(2)}(0)$ in~(a) shows that
    this corresponds to states with~$g^{(2)}\le 1$. It is possible to
    have antibunched mixed states. Four isolines of constant
    populations are shown on both panels. (c) Antibunching and ratio
    of the effective linewidths of the target and source along the
    line I--II, showing a transfer of properties from the source to
    its target. (d) Antibunching and diagonal elements of the density
    matrix along the line III--IV. (e) Density matrices for three
    pumping powers. At~$\Omega_3$, the system is in state~$\ket{1}$
    with more than 60\% probability, although not in a pure state
    anymore.}
  \label{fig:2}
\end{figure}

Concretely, our problem is the dynamic of the system described by
Eq.~(\ref{eq:lunabr27150838CEST2015}) when it is excited by the output
of the system described by Eq.~(\ref{eq:lunabr20123138CEST2015}).
%
The problem of separating the dynamics of coupled systems so that the
source remains oblivious of the effect of its emitted photons onto a
target is tackled in the framework of cascaded
systems~\cite{gardiner_book00a}.  The source and target are dealt with
in the quantum Langevin formalism with input and output fields such
that the output field of the source is, with some delay, the input
field of the target. Imposing causality, one reaches equations of
motion where the source has no dependence on any operators from the
target, which, on the contrary, depends on operators of the source.
From there on, the quantum Langevin equation is converted to a quantum
Ito stochastic form, allowing its expression as a master equation of
the
type~$\partial_t\rho=i[\rho,H]+\sum_{i=1,2}\frac{\gamma_i}{2}\mathcal{L}_{c_i}-\sqrt{\gamma_1\gamma_2}\left\{[\ud{c_1},c_2\rho]+[\rho\ud{c_2},c_1]\right\}$
%
with~$H=H_1+H_2$ the combined Hamiltonian,
$\mathcal{L}_c=(\gamma_c/2)(2c\rho\ud{c}-\rho\ud{c}c-\ud{c}c\rho)$ the
Liouvillian in Lindblad form and $c_2$, $c_1$ the operators from the
source/target subsystems---in our case corresponding to $c_1=\sigma$
and~$c_2=a$---that couple linearly to the output/input fields.  We now
tackle explicit cases of interest.

Fock steady state. --- We first consider the simplest possible
implementation: the excitation of a passive cavity, i.e., an harmonic
oscillator, by the output of a weakly driven two-level system
according to Eq.~(\ref{eq:lunabr20123138CEST2015}). There are only two
parameters ruling this configuration: the ratio of decay rate of the
target with the emission rate of the source, $\gamma_a/\gamma_\sigma$,
and the pumping strength of the source~$\Omega$ (also normalized
to~$\gamma_\sigma$ to keep variables unitless). Figure~\ref{fig:2}
shows the different states of the light field that can be reached in
such a configuration through (a) the photon statistics (color-coded)
and population (isolines) as well as (b) the purity of the state
measured through $\mathrm{Tr}[\rho^2]$ (zero corresponding to
maximally mixed states and one to pure states). The results of the
calculation show that a large family of steady-state pure quantum
states, i.e., with a wavefunction
$\ket{\psi}_\mathrm{ss}=\sum_{i=0}^\infty\sqrt{\rho_{i,i}}\ket{i}$,
can be obtained despite the driven-open nature of the system. Many of
these states are non-classical, sustaining quantum superpositions and
sub-Poissonian fluctuations, depending on the interplay of quantum
pumping and decay.  The cut along the line I--II shows the transition
from a regime where the target behaves according to its own classical
nature (I) to one where it inherits instead the properties of the
quantum source (II). In the former case, where the repetition rate of
the emitter is larger than the decay rate, the many excitations that
can be accumulated give predominance to the target, that grows a
coherent state ($g^{(2)}=1$) and exhibits the same PL spectrum as it
would than if excited classically. In the latter case, on the
contrary, where the input is sparse, the target simply stores the
excitation and reproduces it faithfully. This is the counterpart of
the classically driven quantum dots in the Heitler regime that produce
single-photons with the coherence of the driving
laser~\cite{matthiesen13a}. Interesting scenario lie in between, where
the state fed in the cavity mixes characteristics of both its input
and recipient. New quantum states of the light-field can thus be
created with no need of quantum engineering, merely by exciting the
system with resonance fluorescence. For instance, one can realize
situations with population larger than 1/2 (that of the emitter) or
even larger than unity and still exhibiting antibunching,
$g^{(2)}<1$. Another cut, III--IV, on the isoline of average
population fixed to unity, shows how antibunching increases as the
amplitude of state $\ket{1}$ increases at the expense of vacuum and
state~$\ket{3}$, keeping the amplitude of state~$\ket{2}$ the same.
We find that whenever $g^{(2)}<1/2$, the population is less than
unity.  While this is outside the scope of this text, we have observed
that with sources with an even higher quantum
character~\cite{sanchezmunoz14a}, it is possible to reach pure
steady-state wavefunctions for the driven harmonic oscillator such
that $\rho_{22}>\max(\rho_{11}/2,\rho_{33})$, that is, with less
quantum fluctuations for two particles in the steady-state than are
normally allowed by spontaneous emission, thereby realizing an
effective dissipative Fock state of two particles.

Mollow spectroscopy. --- We now consider another application of
quantum excitation, driving the emitter in the Mollow regime of a
spectral triplet~\cite{mollow69a} and in the presence of interactions
for the target. The Mollow triplet is a treasure trove of photon
correlations when selecting in frequency
windows~\cite{gonzaleztudela13a}. Even though the emitter itself is a
single-photon source, frequency selection allows to access the full
underlying dynamics of the emitter, that is otherwise averaged over to
reduce the physics to a mere antibunching. For instance, while photons
from the side peaks are neatly antibunched, those from the central
peak are slightly bunched, cf.~Fig.~\ref{fig:1}(c). Less prone to
attention, the emission halfway between the central peak and each
satellite is however the most promising for applications as it
involves virtual states. Photons at these frequencies originate from
transitions between every other manifold by ``leapfrog processes''
jumping over an intermediate manifold. This leads to super-bunched
strongly correlated pairs of identical photons, violating
Cauchy-Schwarz and Bell's inequalities~\cite{sanchezmunoz14b}. Photons
in other frequency windows span intermediate cases. This has been
recently confirmed experimentally~\cite{peiris15a}.  Our scheme thus
provides a rich variety of different quantum light, that can be
scanned over the target to probe its response to all types of input,
from single-photon light to super-bunched, strongly-correlated photon
pairs.  
%

\begin{figure}[t]
  \centering
  \includegraphics[width=\linewidth]{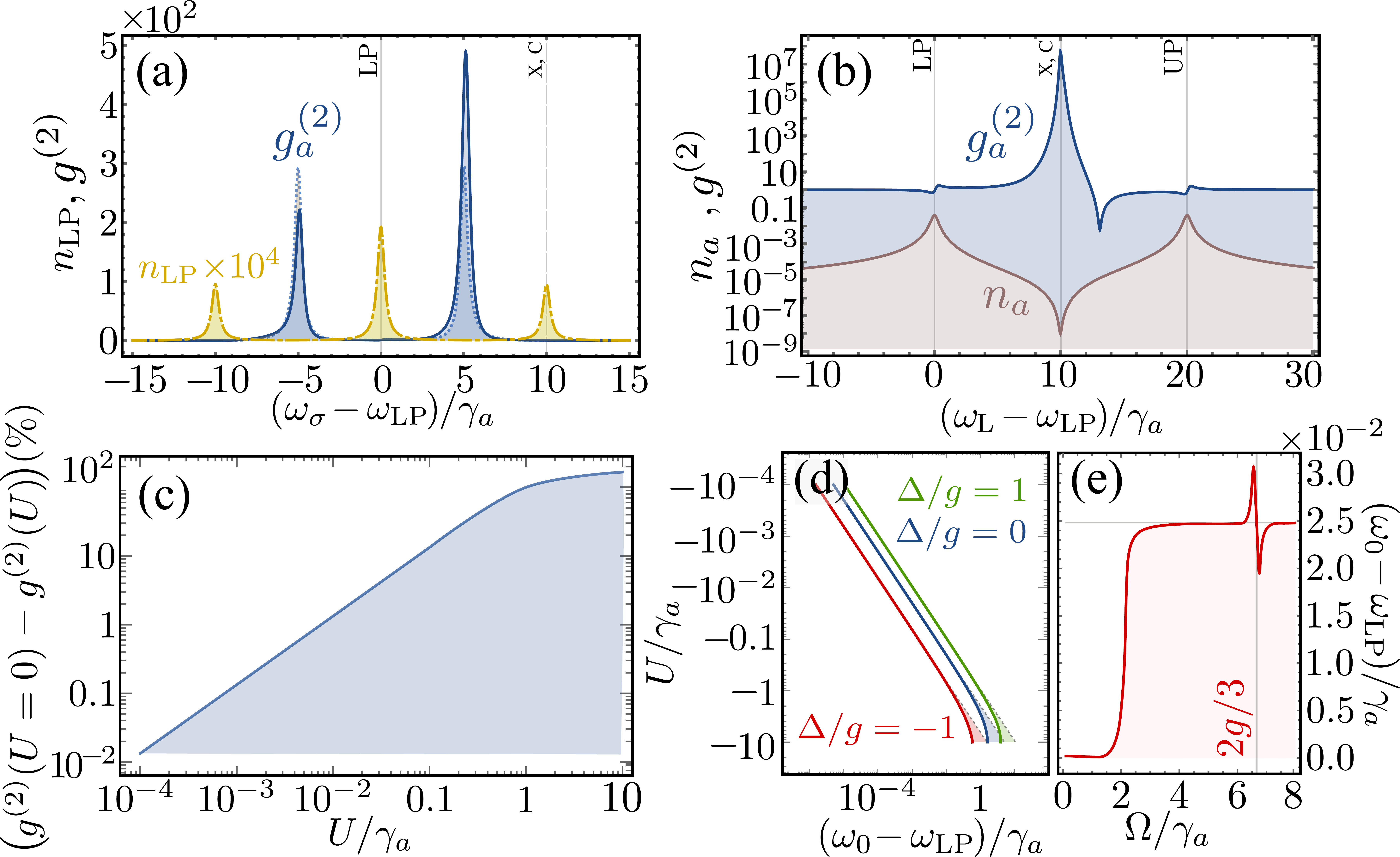}
  \caption{(Colour online) Mollow spectroscopy applied to a
    weakly-interacting Bose gas. (a) Population (dotted-dashed yellow)
    and photon-statistics (solid blue; with $U=0$ in dotted blue) when
    scanning the Mollow triplet onto the lower polariton branch. There
    is an effect on the statistics but not in the population. (b) The
    classical excitation features a weak polariton blockade around the
    lower polariton and a stronger unconventional blockade at higher
    energies, at the cost of diminished signal. Higher pumping washes
    out the effect and lead to a blueshift instead. (c) Magnitude of
    the deviation in photon statistics for a target under Mollow
    pumping as function of $U/\gamma_a$. (d) Value of~$U$ read from
    the shift $\omega_0^*-\omega_\mathrm{LP}$,
    cf.~Eq.~(\ref{eq:lunabr27200758CEST2015}), for various
    detunings~$\Delta$. The dependence is linear up to
    $U\approx\gamma_a$. (e) Mollow splitting required for the
    measurement of~$\omega_0$ to be $\Omega$-independent, allowing an
    absolute measurement of~$U$. There is a resonance at~$\Omega=2g/3$
    when the Mollow satellites meet the two polariton branches.}
  \label{fig:3}
\end{figure}

The responses of weakly-interacting polaritons to classical light (a
laser, right) and to the Mollow triplet (left) are compared on the top
row of Fig.~\ref{fig:3}. Panel~(a) shows the population (dotted-dashed
yellow) and the photon statistics (solid blue) as the quantum source
is scanned over the lower polariton branch, at
energy~$\omega_\mathrm{LP}=(\omega_a+\omega_b)/2-\sqrt{g^2-[(\gamma_a-\gamma_b)/4]^2}$
(we take into account the small shift due to dissipation). In the
population, the triplet is faithfully mapped to the cavity, as
expected. In particular, there is no noticeable shift of the
population (or blueshift in photoluminescence [not shown]). In stark
contrast, the photon statistics 
deviates notably from the linear (non-interacting, $U=0$) case, shown
in a dotted blue line in Fig.~\ref{fig:3}(a). For the classical
excitation, in Panel~(b), the situation is reversed: the weak
nonlinearity requires a strong driving to result in noticeable effects
that manifest better in the population (or photoluminescence), while
the photon statistics, in contrast to the quantum version, remains
close to unity. In both cases, parameters are, in units of $\gamma_a$:
$g=10\gamma_a$, $\gamma_b=0.02\gamma_a$, $\gamma_\sigma=0.01\gamma_a$,
$\Omega=0.5g$ for the quantum excitation and
$\Omega_\mathrm{C}=10^{-2}g$ for the classical one. We considered for
the Figure a large value for~$U/\gamma_a$ of 1/2 to magnify the
observables, in particular displaying neatly both the
conventional~\cite{verger06a} and
unconventional~\cite{liew10a,bamba11a} polariton blockades for the
classical pumping.  The above trend of a strong response in
statistics/small in population to quantum light (and vice versa for
classical excitation) is more marked for smaller---and more realistic
experimentally---value of $U/\gamma_a\ll1$.  Deviations in statistics
for the quantum pumping are shown in Fig.~\ref{fig:3}(c), $\approx
1\%$ for $U/\gamma_a=10^{-2}$. This is a delicate measurement but one
within reach of state of the art experiments~\cite{peiris15a}. 
This is a crucial dissimilitude that powers Mollow spectroscopy.  A
strong pumping of polaritons leads to several complications that
hinder a total or compelling understanding, such as heating,
phase-space filling~\cite{schmittrink85a}, loss of
strong-coupling~\cite{houdre95a}, population of an exciton reservoir
contributing the bulk of the blueshift~\cite{ferrier11a,christmann12a}
and still other factors~\cite{arXiv_dominici13a}.  Since polariton
interactions provide the foundation for nonlinear effects that
constitute much of the polariton literature, the question of their
nature and magnitude could be regarded as one of the most important
open problems of the field~\cite{rossbach13a,cilibrizzi14a}.  In
contrast, the excitation of the same system with the Mollow triplet is
clean as it avoids such complications of high densities.  Since it
recourses to the minimum amount of polaritons required to poke the
interaction (two), Mollow spectroscopy acts as a ``probe'' in the
ultimate sense of the term, with as little disturbance as
possible. The fact that polaritons are consistently probed in pairs
not only optimizes the effect of the interaction, it also allows along
with the tunability of statistics from the Mollow triplet to extract
the numerical value of the nonlinearity. This is achieved by measuring
the change in photon statistics with the frequency of excitation. The
maximum super-bunching is provided by the leapfrog photon pairs,
emitted in the two frequency windows between the central peak and the
satellites. Photons closer to the satellites lose this super-bunching
faster than those closer to the central peak.  Comparing the response
of the system when going in these two directions allows to quantify
small nonlinearities, otherwise hidden in the radiative broadening.
Thanks to the symmetry of the Mollow triplet, such a comparison can be
conveniently implemented without the need for a calibration of the
frequencies, by using directly both sides of the triplet, rather than
both sides of the super-bunching peak. One can simply sweep the Mollow
lineshape onto the target and records its photon statistics. From
these measurements, one then defines~$f$ the auto-convolution of the
correlation function, $f=g^{(2)}\ast g^{(2)}$, i.e.,
$f(\omega_0)=\int_{-\infty}^{+\infty}g^{(2)}(\omega_0-\omega)g^{(2)}(\omega)\,d\omega$. The
triangle inequality places the maximum of~$f(\omega)$ at the
value~$\omega_0^*$ that minimizes the asymmetry of $g^{(2)}(\omega)$
around~$\omega_0-\omega_\mathrm{LP}$. When the leapfrog processes are
sufficiently well defined (see below), this gives precisely the shift
caused by the two-polariton interaction:
\begin{equation}
  \label{eq:lunabr27200758CEST2015}
  \omega_0^*=\omega_\mathrm{LP}-2 U\chi_{02}^2\,,
\end{equation}
where~$\chi_{02}$ is the two-polariton Hopfield coefficient for the
state~$\ket{02}$ of two excitons and zero photon. The closed-form
expression of~$\chi_{02}$ in terms of detuning is too cumbersome to be
given here but the value is straightforwardly obtained by
diagonalization of~Eq.~(\ref{eq:lunabr27150838CEST2015}). While the
interaction affects~$\chi_{02}$ in principle, for most of the range of
interest, the effect is completely negligible, as seen in
Fig.~\ref{fig:3}(d). For instance, at resonance,
$\chi_{02}^2=1/4$. More importantly, there is no dependence on the
population or other dynamical variables and the measurement is thus
absolute, unlike the blueshift from a classical driving that, on the
contrary, requires a careful calibration of the laser
intensity. 

If the Mollow triplet is not neatly formed, the leapfrog processes are
mixed with other types of less correlated emission, leading to some
departure from a shift ruled wholly by the two-polariton
interaction. Figure~\ref{fig:3}(e) shows the splitting required for
Eq.~(\ref{eq:lunabr27200758CEST2015}) to be accurate. Even at low
splitting, from a power dependence, it is however possible to estimate
the nonlinearity. If large Mollow splittings as compared to the Rabi
splitting are available, a resonance in~$g^{(2)}$ is obtained at
$\Omega=3g/2$ when the lower leapfrog excite the lower polariton
branch while the upper Mollow satellite excites the upper polariton
branch, as shown in Fig.~\ref{fig:3}(e).


Conclusions and Perspectives. --- We have shown how exciting a system
with quantum light opens new perspective in several areas of quantum
physics. Exciting a passive system, we have shown how to realize
various pure quantum states in the steady state. Exciting interacting
ones, we have shown a deep change of paradigm when trading the
classical excitation for a quantum one. In the latter case, one has a
strong response in the correlated emission from the weakly interacting
polaritons even with small populations while in the other case, one
has to recourse to a strong excitation for the interaction to play a
role, which they do in the population but at the expense of the
quantum correlations that are washed out in the macroscopically
occupied state driven by the laser. This allows us to propose a new
spectroscopic technique, extracting system parameters otherwise
unreachable due to the intensity required to have them play a role
and/or overcome radiative broadening.

These findings open multiple avenues along similar lines: another
direct application of the same technique is to use the polarization
degree of freedom to measure the spin-dependent interaction.
A more involved setup that allows to diffract the incoming Mollow beam
into two beams with wavevectors~$\mathbf{k_1}$ and~$\mathbf{k_2}$,
keeping their frequency otherwise the same, should similarly allow to
read from shifts in cross-correlated $g^{(2)}$ the value of the
exchanged-momentum matrix element~$U(\mathbf{q})$ in the full
polariton Hamiltonian. This Hamiltonian features the coupling term
$\sum_{\mathbf{k}_1,\mathbf{k}_2,\mathbf{q}}U_{\mathbf{k}_1\mathbf{k}_2}(\mathbf{q})\ud{b_{\mathbf{k}_1+\mathbf{q}}}\ud{b_{\mathbf{k}_2-\mathbf{q}}}b_{\mathbf{k}_1}b_{\mathbf{k}_2}$,
of which we have just worked out in detail the
case~$\mathbf{q}=0$. Spanning over~$\mathbf{q}$ will allow an accurate
and comprehensive reconstruction of the full interaction potential,
also clean from saturations effects, fluctuations and allowing to
detect even minute effects in strongly dissipative environments,
making available spectroscopic tools that should allow to revisit the
precious few results that exist on the dispersion of excitations of
polariton gases~\cite{utsunomiya08a,kohnle11a}. These results go
beyond the scope of this Letter and will be presented separately.  We
have focused here on weakly linear systems, in fact even showing
nontrivial results with a completely linear system.  The same
formalism can be applied with equal ease to targets that are also
strongly quantum, such as, but not exclusively, a Jaynes--Cummings
system, a chain or lattice of two-level systems, an optical-parametric
oscillator, etc. The source itself is not limited to the Mollow
triplet. The emitter of Fock states recently proposed by some of the
Authors~\cite{sanchezmunoz14a}, releasing all its energy in bundles
of~$N$ photons, will be a key resource for quantum excitation. Beyond
the exotic states of light already alluded to, this should open the
door to new classes of excitations, such as the correlated
electron-hole clusters, the so-called
dropletons~\cite{almandhunter14a}, recently discovered within the
limitations of classical excitations thanks to a feast of theoretical
deduction.

\section{Acknowledgements}
We thank Amir Rahmani and Kai M\"uller for discussions.  We
acknowledge funding by the Spanish MINECO (FPI \& RyC programs) and
the EU under the ERC scheme POLAFLOW.
\bibliography{Sci,books,arXiv}

\begin{thebibliography}{40}
\expandafter\ifx\csname natexlab\endcsname\relax\def\natexlab#1{#1}\fi
\expandafter\ifx\csname bibnamefont\endcsname\relax
  \def\bibnamefont#1{#1}\fi
\expandafter\ifx\csname bibfnamefont\endcsname\relax
  \def\bibfnamefont#1{#1}\fi
\expandafter\ifx\csname citenamefont\endcsname\relax
  \def\citenamefont#1{#1}\fi
\expandafter\ifx\csname url\endcsname\relax
  \def\url#1{\texttt{#1}}\fi
\expandafter\ifx\csname urlprefix\endcsname\relax\def\urlprefix{URL }\fi
\providecommand{\bibinfo}[2]{#2}
\providecommand{\eprint}[2][]{\url{#2}}

\bibitem[{\citenamefont{O'Brien et~al.}(2009)\citenamefont{O'Brien, Furusawa,
  and Vuckovic}}]{obrien09a}
\bibinfo{author}{\bibfnamefont{J.~L.} \bibnamefont{O'Brien}},
  \bibinfo{author}{\bibfnamefont{A.}~\bibnamefont{Furusawa}}, \bibnamefont{and}
  \bibinfo{author}{\bibfnamefont{J.}~\bibnamefont{Vuckovic}},
  \bibinfo{journal}{Nat. Phys.} \textbf{\bibinfo{volume}{3}},
  \bibinfo{pages}{687} (\bibinfo{year}{2009}).

\bibitem[{\citenamefont{Bertolotti et~al.}(2015)\citenamefont{Bertolotti,
  Bovino, and Sibilia}}]{bertolotti15a}
\bibinfo{author}{\bibfnamefont{M.}~\bibnamefont{Bertolotti}},
  \bibinfo{author}{\bibfnamefont{F.}~\bibnamefont{Bovino}}, \bibnamefont{and}
  \bibinfo{author}{\bibfnamefont{C.}~\bibnamefont{Sibilia}},
  \bibinfo{journal}{Progress in Optics} p. \bibinfo{pages}{In Press}
  (\bibinfo{year}{2015}).

\bibitem[{\citenamefont{Dodonov}(2002)}]{dodonov02a}
\bibinfo{author}{\bibfnamefont{V.~V.} \bibnamefont{Dodonov}},
  \bibinfo{journal}{J. Opt. B} \textbf{\bibinfo{volume}{4}},
  \bibinfo{pages}{R1} (\bibinfo{year}{2002}).

\bibitem[{\citenamefont{M\"uller et~al.}(2015)\citenamefont{M\"uller,
  Rundquist, Fischer, Sarmiento, Lagoudakis, Kelaita, noz, del Valle, Laussy,
  and Vuckovic}}]{muller15a}
\bibinfo{author}{\bibfnamefont{K.}~\bibnamefont{M\"uller}},
  \bibinfo{author}{\bibfnamefont{A.}~\bibnamefont{Rundquist}},
  \bibinfo{author}{\bibfnamefont{K.~A.} \bibnamefont{Fischer}},
  \bibinfo{author}{\bibfnamefont{T.}~\bibnamefont{Sarmiento}},
  \bibinfo{author}{\bibfnamefont{K.~G.} \bibnamefont{Lagoudakis}},
  \bibinfo{author}{\bibfnamefont{Y.~A.} \bibnamefont{Kelaita}},
  \bibinfo{author}{\bibfnamefont{C.~S.~M.} \bibnamefont{noz}},
  \bibinfo{author}{\bibfnamefont{E.}~\bibnamefont{del Valle}},
  \bibinfo{author}{\bibfnamefont{F.~P.} \bibnamefont{Laussy}},
  \bibnamefont{and} \bibinfo{author}{\bibfnamefont{J.}~\bibnamefont{Vuckovic}},
  \bibinfo{journal}{Phys. Rev. Lett.}  (\bibinfo{year}{2015}).

\bibitem[{\citenamefont{{Javier Garc\'ia de
  Abajo}}(2012)}]{javiergarciadeabajo12a}
\bibinfo{author}{\bibfnamefont{F.}~\bibnamefont{{Javier Garc\'ia de Abajo}}},
  \bibinfo{journal}{Nature} \textbf{\bibinfo{volume}{483}},
  \bibinfo{pages}{417} (\bibinfo{year}{2012}).

\bibitem[{\citenamefont{Kavokin et~al.}(2011)\citenamefont{Kavokin, Baumberg,
  Malpuech, and Laussy}}]{kavokin_book11a}
\bibinfo{author}{\bibfnamefont{A.}~\bibnamefont{Kavokin}},
  \bibinfo{author}{\bibfnamefont{J.~J.} \bibnamefont{Baumberg}},
  \bibinfo{author}{\bibfnamefont{G.}~\bibnamefont{Malpuech}}, \bibnamefont{and}
  \bibinfo{author}{\bibfnamefont{F.~P.} \bibnamefont{Laussy}},
  \emph{\bibinfo{title}{Microcavities}} (\bibinfo{publisher}{Oxford University
  Press}, \bibinfo{year}{2011}), \bibinfo{edition}{2nd} ed.

\bibitem[{\citenamefont{Verger et~al.}(2006)\citenamefont{Verger, Ciuti, and
  Carusotto}}]{verger06a}
\bibinfo{author}{\bibfnamefont{A.}~\bibnamefont{Verger}},
  \bibinfo{author}{\bibfnamefont{C.}~\bibnamefont{Ciuti}}, \bibnamefont{and}
  \bibinfo{author}{\bibfnamefont{I.}~\bibnamefont{Carusotto}},
  \bibinfo{journal}{Phys. Rev. B} \textbf{\bibinfo{volume}{73}},
  \bibinfo{pages}{193306} (\bibinfo{year}{2006}).

\bibitem[{\citenamefont{Boulier et~al.}(2014)\citenamefont{Boulier, Bamba, Amo,
  Adrados, Lemaitre, Galopin, Sagnes, Bloch, Ciuti, Giacobino
  et~al.}}]{boulier14a}
\bibinfo{author}{\bibfnamefont{T.}~\bibnamefont{Boulier}},
  \bibinfo{author}{\bibfnamefont{M.}~\bibnamefont{Bamba}},
  \bibinfo{author}{\bibfnamefont{A.}~\bibnamefont{Amo}},
  \bibinfo{author}{\bibfnamefont{C.}~\bibnamefont{Adrados}},
  \bibinfo{author}{\bibfnamefont{A.}~\bibnamefont{Lemaitre}},
  \bibinfo{author}{\bibfnamefont{E.}~\bibnamefont{Galopin}},
  \bibinfo{author}{\bibfnamefont{I.}~\bibnamefont{Sagnes}},
  \bibinfo{author}{\bibfnamefont{J.}~\bibnamefont{Bloch}},
  \bibinfo{author}{\bibfnamefont{C.}~\bibnamefont{Ciuti}},
  \bibinfo{author}{\bibfnamefont{E.}~\bibnamefont{Giacobino}},
  \bibnamefont{et~al.}, \bibinfo{journal}{Nat. Comm.}
  \textbf{\bibinfo{volume}{5}} (\bibinfo{year}{2014}).

\bibitem[{\citenamefont{Kasprzak et~al.}(2006)\citenamefont{Kasprzak, Richard,
  Kundermann, Baas, Jeambrun, Keeling, Marchetti, Szymanska, Andr\'e, Staehli
  et~al.}}]{kasprzak06a}
\bibinfo{author}{\bibfnamefont{J.}~\bibnamefont{Kasprzak}},
  \bibinfo{author}{\bibfnamefont{M.}~\bibnamefont{Richard}},
  \bibinfo{author}{\bibfnamefont{S.}~\bibnamefont{Kundermann}},
  \bibinfo{author}{\bibfnamefont{A.}~\bibnamefont{Baas}},
  \bibinfo{author}{\bibfnamefont{P.}~\bibnamefont{Jeambrun}},
  \bibinfo{author}{\bibfnamefont{J.~M.~J.} \bibnamefont{Keeling}},
  \bibinfo{author}{\bibfnamefont{F.~M.} \bibnamefont{Marchetti}},
  \bibinfo{author}{\bibfnamefont{M.~H.} \bibnamefont{Szymanska}},
  \bibinfo{author}{\bibfnamefont{R.}~\bibnamefont{Andr\'e}},
  \bibinfo{author}{\bibfnamefont{J.~L.} \bibnamefont{Staehli}},
  \bibnamefont{et~al.}, \bibinfo{journal}{Nature}
  \textbf{\bibinfo{volume}{443}}, \bibinfo{pages}{409} (\bibinfo{year}{2006}).

\bibitem[{\citenamefont{Amo et~al.}(2009)\citenamefont{Amo, Lefr\`ere, Pigeon,
  Adrados, Ciuti, Carusotto, Houdr\'e, Giacobino, and Bramati}}]{amo09b}
\bibinfo{author}{\bibfnamefont{A.}~\bibnamefont{Amo}},
  \bibinfo{author}{\bibfnamefont{J.}~\bibnamefont{Lefr\`ere}},
  \bibinfo{author}{\bibfnamefont{S.}~\bibnamefont{Pigeon}},
  \bibinfo{author}{\bibfnamefont{C.}~\bibnamefont{Adrados}},
  \bibinfo{author}{\bibfnamefont{C.}~\bibnamefont{Ciuti}},
  \bibinfo{author}{\bibfnamefont{I.}~\bibnamefont{Carusotto}},
  \bibinfo{author}{\bibfnamefont{R.}~\bibnamefont{Houdr\'e}},
  \bibinfo{author}{\bibfnamefont{E.}~\bibnamefont{Giacobino}},
  \bibnamefont{and} \bibinfo{author}{\bibfnamefont{A.}~\bibnamefont{Bramati}},
  \bibinfo{journal}{Nat. Phys.} \textbf{\bibinfo{volume}{5}},
  \bibinfo{pages}{805} (\bibinfo{year}{2009}).

\bibitem[{\citenamefont{Lagoudakis et~al.}(2010)\citenamefont{Lagoudakis,
  Pietka, Wouters, André, and Deveaud-Plédran}}]{lagoudakis10a}
\bibinfo{author}{\bibfnamefont{K.~G.} \bibnamefont{Lagoudakis}},
  \bibinfo{author}{\bibfnamefont{B.}~\bibnamefont{Pietka}},
  \bibinfo{author}{\bibfnamefont{M.}~\bibnamefont{Wouters}},
  \bibinfo{author}{\bibfnamefont{R.}~\bibnamefont{André}}, \bibnamefont{and}
  \bibinfo{author}{\bibfnamefont{B.}~\bibnamefont{Deveaud-Plédran}},
  \bibinfo{journal}{Phys. Rev. Lett.} \textbf{\bibinfo{volume}{105}},
  \bibinfo{pages}{120403} (\bibinfo{year}{2010}).

\bibitem[{\citenamefont{Abbarchi et~al.}(2013)\citenamefont{Abbarchi, Amo,
  Sala, Solnyshkov, Flayac, Ferrier, Sagnes, Galopin, Lema\^itre, Malpuech
  et~al.}}]{abbarchi13a}
\bibinfo{author}{\bibfnamefont{M.}~\bibnamefont{Abbarchi}},
  \bibinfo{author}{\bibfnamefont{A.}~\bibnamefont{Amo}},
  \bibinfo{author}{\bibfnamefont{V.~G.} \bibnamefont{Sala}},
  \bibinfo{author}{\bibfnamefont{D.~D.} \bibnamefont{Solnyshkov}},
  \bibinfo{author}{\bibfnamefont{H.}~\bibnamefont{Flayac}},
  \bibinfo{author}{\bibfnamefont{L.}~\bibnamefont{Ferrier}},
  \bibinfo{author}{\bibfnamefont{I.}~\bibnamefont{Sagnes}},
  \bibinfo{author}{\bibfnamefont{E.}~\bibnamefont{Galopin}},
  \bibinfo{author}{\bibfnamefont{A.}~\bibnamefont{Lema\^itre}},
  \bibinfo{author}{\bibfnamefont{G.}~\bibnamefont{Malpuech}},
  \bibnamefont{et~al.}, \bibinfo{journal}{Nat. Phys.}
  \textbf{\bibinfo{volume}{9}}, \bibinfo{pages}{275} (\bibinfo{year}{2013}).

\bibitem[{\citenamefont{Byrnes et~al.}(2011)\citenamefont{Byrnes, Yan, and
  Yamamoto}}]{byrnes11a}
\bibinfo{author}{\bibfnamefont{T.}~\bibnamefont{Byrnes}},
  \bibinfo{author}{\bibfnamefont{K.}~\bibnamefont{Yan}}, \bibnamefont{and}
  \bibinfo{author}{\bibfnamefont{Y.}~\bibnamefont{Yamamoto}},
  \bibinfo{journal}{New J. Phys.} \textbf{\bibinfo{volume}{13}},
  \bibinfo{pages}{113025} (\bibinfo{year}{2011}).

\bibitem[{\citenamefont{Byrnes et~al.}(2013)\citenamefont{Byrnes, Koyama, K,
  and Yamamoto}}]{byrnes13a}
\bibinfo{author}{\bibfnamefont{T.}~\bibnamefont{Byrnes}},
  \bibinfo{author}{\bibfnamefont{S.}~\bibnamefont{Koyama}},
  \bibinfo{author}{\bibfnamefont{K.~Y.} \bibnamefont{K}}, \bibnamefont{and}
  \bibinfo{author}{\bibfnamefont{Y.}~\bibnamefont{Yamamoto}},
  \bibinfo{journal}{Scientific Report} \textbf{\bibinfo{volume}{3}},
  \bibinfo{pages}{2531} (\bibinfo{year}{2013}).

\bibitem[{\citenamefont{Ballarini et~al.}(2013)\citenamefont{Ballarini, Giorgi,
  Cancellieri, Houdr\'e, Giacobino, Cingolani, Bramati, Gigli, and
  Sanvitto}}]{ballarini13a}
\bibinfo{author}{\bibfnamefont{D.}~\bibnamefont{Ballarini}},
  \bibinfo{author}{\bibfnamefont{M.~D.} \bibnamefont{Giorgi}},
  \bibinfo{author}{\bibfnamefont{E.}~\bibnamefont{Cancellieri}},
  \bibinfo{author}{\bibfnamefont{R.}~\bibnamefont{Houdr\'e}},
  \bibinfo{author}{\bibfnamefont{E.}~\bibnamefont{Giacobino}},
  \bibinfo{author}{\bibfnamefont{R.}~\bibnamefont{Cingolani}},
  \bibinfo{author}{\bibfnamefont{A.}~\bibnamefont{Bramati}},
  \bibinfo{author}{\bibfnamefont{G.}~\bibnamefont{Gigli}}, \bibnamefont{and}
  \bibinfo{author}{\bibfnamefont{D.}~\bibnamefont{Sanvitto}},
  \bibinfo{journal}{Nat. Comm.} \textbf{\bibinfo{volume}{4}},
  \bibinfo{pages}{1778} (\bibinfo{year}{2013}).

\bibitem[{\citenamefont{Ant\'on et~al.}(2013)\citenamefont{Ant\'on, Liew,
  Cuadra, Mart\'in, Eldridge, Hatzopoulos, Stavrinidis, Savvidis, and
  na}}]{anton13a}
\bibinfo{author}{\bibfnamefont{C.}~\bibnamefont{Ant\'on}},
  \bibinfo{author}{\bibfnamefont{T.~C.~H.} \bibnamefont{Liew}},
  \bibinfo{author}{\bibfnamefont{J.}~\bibnamefont{Cuadra}},
  \bibinfo{author}{\bibfnamefont{M.~D.} \bibnamefont{Mart\'in}},
  \bibinfo{author}{\bibfnamefont{P.~S.} \bibnamefont{Eldridge}},
  \bibinfo{author}{\bibfnamefont{Z.}~\bibnamefont{Hatzopoulos}},
  \bibinfo{author}{\bibfnamefont{G.}~\bibnamefont{Stavrinidis}},
  \bibinfo{author}{\bibfnamefont{P.~G.} \bibnamefont{Savvidis}},
  \bibnamefont{and} \bibinfo{author}{\bibfnamefont{L.~V.} \bibnamefont{na}},
  \bibinfo{journal}{Phys. Rev. B} \textbf{\bibinfo{volume}{88}},
  \bibinfo{pages}{245307} (\bibinfo{year}{2013}).

\bibitem[{\citenamefont{Kira and Koch}(2006)}]{kira06a}
\bibinfo{author}{\bibfnamefont{M.}~\bibnamefont{Kira}} \bibnamefont{and}
  \bibinfo{author}{\bibfnamefont{S.~W.} \bibnamefont{Koch}},
  \bibinfo{journal}{Phys. Rev. A} \textbf{\bibinfo{volume}{73}},
  \bibinfo{pages}{013813} (\bibinfo{year}{2006}).

\bibitem[{\citenamefont{Kira et~al.}(2011)\citenamefont{Kira, Koch, Smith,
  Hunter, and Cundiff}}]{kira11a}
\bibinfo{author}{\bibfnamefont{M.}~\bibnamefont{Kira}},
  \bibinfo{author}{\bibfnamefont{S.~W.} \bibnamefont{Koch}},
  \bibinfo{author}{\bibfnamefont{R.~P.} \bibnamefont{Smith}},
  \bibinfo{author}{\bibfnamefont{A.~E.} \bibnamefont{Hunter}},
  \bibnamefont{and} \bibinfo{author}{\bibfnamefont{S.~T.}
  \bibnamefont{Cundiff}}, \bibinfo{journal}{Nat. Phys.}
  \textbf{\bibinfo{volume}{7}}, \bibinfo{pages}{799} (\bibinfo{year}{2011}).

\bibitem[{\citenamefont{Carmele et~al.}(2009)\citenamefont{Carmele, Knorr, and
  Richter}}]{carmele09a}
\bibinfo{author}{\bibfnamefont{A.}~\bibnamefont{Carmele}},
  \bibinfo{author}{\bibfnamefont{A.}~\bibnamefont{Knorr}}, \bibnamefont{and}
  \bibinfo{author}{\bibfnamefont{M.}~\bibnamefont{Richter}},
  \bibinfo{journal}{Phys. Rev. B} \textbf{\bibinfo{volume}{79}},
  \bibinfo{pages}{035316} (\bibinfo{year}{2009}).

\bibitem[{\citenamefont{A{\ss}mann and Bayer}(2011)}]{assmann11b}
\bibinfo{author}{\bibfnamefont{M.}~\bibnamefont{A{\ss}mann}} \bibnamefont{and}
  \bibinfo{author}{\bibfnamefont{M.}~\bibnamefont{Bayer}},
  \bibinfo{journal}{Phys. Rev. A} \textbf{\bibinfo{volume}{84}},
  \bibinfo{pages}{053806} (\bibinfo{year}{2011}).

\bibitem[{\citenamefont{Mollow}(1969)}]{mollow69a}
\bibinfo{author}{\bibfnamefont{B.~R.} \bibnamefont{Mollow}},
  \bibinfo{journal}{Phys. Rev.} \textbf{\bibinfo{volume}{188}},
  \bibinfo{pages}{1969} (\bibinfo{year}{1969}).

\bibitem[{\citenamefont{del Valle and Laussy}(2010)}]{delvalle10d}
\bibinfo{author}{\bibfnamefont{E.}~\bibnamefont{del Valle}} \bibnamefont{and}
  \bibinfo{author}{\bibfnamefont{F.~P.} \bibnamefont{Laussy}},
  \bibinfo{journal}{Phys. Rev. Lett.} \textbf{\bibinfo{volume}{105}},
  \bibinfo{pages}{233601} (\bibinfo{year}{2010}).

\bibitem[{\citenamefont{Gardiner and Zoller}(2000)}]{gardiner_book00a}
\bibinfo{author}{\bibfnamefont{G.~W.} \bibnamefont{Gardiner}} \bibnamefont{and}
  \bibinfo{author}{\bibfnamefont{P.}~\bibnamefont{Zoller}},
  \emph{\bibinfo{title}{Quantum Noise}} (\bibinfo{publisher}{Springer-Verlag,
  Berlin}, \bibinfo{year}{2000}), \bibinfo{edition}{2nd} ed.

\bibitem[{\citenamefont{Matthiesen et~al.}(2013)\citenamefont{Matthiesen,
  Geller, Schulte, Gall, Hansom, Li, Hugues, Clarke, and
  Atat\"ure}}]{matthiesen13a}
\bibinfo{author}{\bibfnamefont{C.}~\bibnamefont{Matthiesen}},
  \bibinfo{author}{\bibfnamefont{M.}~\bibnamefont{Geller}},
  \bibinfo{author}{\bibfnamefont{C.~H.~H.} \bibnamefont{Schulte}},
  \bibinfo{author}{\bibfnamefont{C.~L.} \bibnamefont{Gall}},
  \bibinfo{author}{\bibfnamefont{J.}~\bibnamefont{Hansom}},
  \bibinfo{author}{\bibfnamefont{Z.}~\bibnamefont{Li}},
  \bibinfo{author}{\bibfnamefont{M.}~\bibnamefont{Hugues}},
  \bibinfo{author}{\bibfnamefont{E.}~\bibnamefont{Clarke}}, \bibnamefont{and}
  \bibinfo{author}{\bibfnamefont{M.}~\bibnamefont{Atat\"ure}},
  \bibinfo{journal}{Nat. Comm.} \textbf{\bibinfo{volume}{4}},
  \bibinfo{pages}{1600} (\bibinfo{year}{2013}).

\bibitem[{\citenamefont{{Sanchez Mu\~noz}
  et~al.}(2014{\natexlab{a}})\citenamefont{{Sanchez Mu\~noz}, del Valle,
  Tudela, M\"uller, Lichtmannecker, Kaniber, Tejedor, Finley, and
  Laussy}}]{sanchezmunoz14a}
\bibinfo{author}{\bibfnamefont{C.}~\bibnamefont{{Sanchez Mu\~noz}}},
  \bibinfo{author}{\bibfnamefont{E.}~\bibnamefont{del Valle}},
  \bibinfo{author}{\bibfnamefont{A.~G.} \bibnamefont{Tudela}},
  \bibinfo{author}{\bibfnamefont{K.}~\bibnamefont{M\"uller}},
  \bibinfo{author}{\bibfnamefont{S.}~\bibnamefont{Lichtmannecker}},
  \bibinfo{author}{\bibfnamefont{M.}~\bibnamefont{Kaniber}},
  \bibinfo{author}{\bibfnamefont{C.}~\bibnamefont{Tejedor}},
  \bibinfo{author}{\bibfnamefont{J.}~\bibnamefont{Finley}}, \bibnamefont{and}
  \bibinfo{author}{\bibfnamefont{F.}~\bibnamefont{Laussy}},
  \bibinfo{journal}{Nat. Photon.} \textbf{\bibinfo{volume}{8}},
  \bibinfo{pages}{550} (\bibinfo{year}{2014}{\natexlab{a}}).

\bibitem[{\citenamefont{Gonzalez-Tudela
  et~al.}(2013)\citenamefont{Gonzalez-Tudela, Laussy, Tejedor, Hartmann, and
  del Valle}}]{gonzaleztudela13a}
\bibinfo{author}{\bibfnamefont{A.}~\bibnamefont{Gonzalez-Tudela}},
  \bibinfo{author}{\bibfnamefont{F.~P.} \bibnamefont{Laussy}},
  \bibinfo{author}{\bibfnamefont{C.}~\bibnamefont{Tejedor}},
  \bibinfo{author}{\bibfnamefont{M.~J.} \bibnamefont{Hartmann}},
  \bibnamefont{and} \bibinfo{author}{\bibfnamefont{E.}~\bibnamefont{del
  Valle}}, \bibinfo{journal}{New J. Phys.} \textbf{\bibinfo{volume}{15}},
  \bibinfo{pages}{033036} (\bibinfo{year}{2013}).

\bibitem[{\citenamefont{{Sanchez Mu\~noz}
  et~al.}(2014{\natexlab{b}})\citenamefont{{Sanchez Mu\~noz}, del Valle,
  Tejedor, and Laussy}}]{sanchezmunoz14b}
\bibinfo{author}{\bibfnamefont{C.}~\bibnamefont{{Sanchez Mu\~noz}}},
  \bibinfo{author}{\bibfnamefont{E.}~\bibnamefont{del Valle}},
  \bibinfo{author}{\bibfnamefont{C.}~\bibnamefont{Tejedor}}, \bibnamefont{and}
  \bibinfo{author}{\bibfnamefont{F.}~\bibnamefont{Laussy}},
  \bibinfo{journal}{Phys. Rev. A} \textbf{\bibinfo{volume}{90}},
  \bibinfo{pages}{052111} (\bibinfo{year}{2014}{\natexlab{b}}).

\bibitem[{\citenamefont{Peiris et~al.}(2015)\citenamefont{Peiris, Petrak,
  Konthasinghe, Yu, Niu, and Muller}}]{peiris15a}
\bibinfo{author}{\bibfnamefont{M.}~\bibnamefont{Peiris}},
  \bibinfo{author}{\bibfnamefont{B.}~\bibnamefont{Petrak}},
  \bibinfo{author}{\bibfnamefont{K.}~\bibnamefont{Konthasinghe}},
  \bibinfo{author}{\bibfnamefont{Y.}~\bibnamefont{Yu}},
  \bibinfo{author}{\bibfnamefont{Z.~C.} \bibnamefont{Niu}}, \bibnamefont{and}
  \bibinfo{author}{\bibfnamefont{A.}~\bibnamefont{Muller}},
  \bibinfo{journal}{Phys. Rev. B} \textbf{\bibinfo{volume}{91}},
  \bibinfo{pages}{195125} (\bibinfo{year}{2015}).

\bibitem[{\citenamefont{Liew and Savona}(2010)}]{liew10a}
\bibinfo{author}{\bibfnamefont{T.~C.~H.} \bibnamefont{Liew}} \bibnamefont{and}
  \bibinfo{author}{\bibfnamefont{V.}~\bibnamefont{Savona}},
  \bibinfo{journal}{Phys. Rev. Lett.} \textbf{\bibinfo{volume}{104}},
  \bibinfo{pages}{183601} (\bibinfo{year}{2010}).

\bibitem[{\citenamefont{Bamba et~al.}(2011)\citenamefont{Bamba, \Imamoglu,
  Carusotto, and Ciuti}}]{bamba11a}
\bibinfo{author}{\bibfnamefont{M.}~\bibnamefont{Bamba}},
  \bibinfo{author}{\bibfnamefont{A.}~\bibnamefont{\Imamoglu}},
  \bibinfo{author}{\bibfnamefont{I.}~\bibnamefont{Carusotto}},
  \bibnamefont{and} \bibinfo{author}{\bibfnamefont{C.}~\bibnamefont{Ciuti}},
  \bibinfo{journal}{Phys. Rev. A} \textbf{\bibinfo{volume}{83}},
  \bibinfo{pages}{021802(R)} (\bibinfo{year}{2011}).

\bibitem[{\citenamefont{Schmitt-Rink et~al.}(1985)\citenamefont{Schmitt-Rink,
  Chemla, and Miller}}]{schmittrink85a}
\bibinfo{author}{\bibfnamefont{S.}~\bibnamefont{Schmitt-Rink}},
  \bibinfo{author}{\bibfnamefont{D.~S.} \bibnamefont{Chemla}},
  \bibnamefont{and} \bibinfo{author}{\bibfnamefont{D.~A.~B.}
  \bibnamefont{Miller}}, \bibinfo{journal}{Phys. Rev. B}
  \textbf{\bibinfo{volume}{32}}, \bibinfo{pages}{6601} (\bibinfo{year}{1985}).

\bibitem[{\citenamefont{Houdr\'e et~al.}(1995)\citenamefont{Houdr\'e, Gibernon,
  Pellandini, Stanley, Oesterle, Weisbuch, O'Gorman, Roycroft, and
  Ilegems}}]{houdre95a}
\bibinfo{author}{\bibfnamefont{R.}~\bibnamefont{Houdr\'e}},
  \bibinfo{author}{\bibfnamefont{J.~L.} \bibnamefont{Gibernon}},
  \bibinfo{author}{\bibfnamefont{P.}~\bibnamefont{Pellandini}},
  \bibinfo{author}{\bibfnamefont{R.~P.} \bibnamefont{Stanley}},
  \bibinfo{author}{\bibfnamefont{U.}~\bibnamefont{Oesterle}},
  \bibinfo{author}{\bibfnamefont{C.}~\bibnamefont{Weisbuch}},
  \bibinfo{author}{\bibfnamefont{J.}~\bibnamefont{O'Gorman}},
  \bibinfo{author}{\bibfnamefont{B.}~\bibnamefont{Roycroft}}, \bibnamefont{and}
  \bibinfo{author}{\bibfnamefont{M.}~\bibnamefont{Ilegems}},
  \bibinfo{journal}{Phys. Rev. B} \textbf{\bibinfo{volume}{52}},
  \bibinfo{pages}{7810} (\bibinfo{year}{1995}).

\bibitem[{\citenamefont{Ferrier et~al.}(2011)\citenamefont{Ferrier, Wertz,
  Johne, Solnyshkov, Senellart, Sagnes, Lema\^itre, Malpuech, and
  Bloch}}]{ferrier11a}
\bibinfo{author}{\bibfnamefont{L.}~\bibnamefont{Ferrier}},
  \bibinfo{author}{\bibfnamefont{E.}~\bibnamefont{Wertz}},
  \bibinfo{author}{\bibfnamefont{R.}~\bibnamefont{Johne}},
  \bibinfo{author}{\bibfnamefont{D.~D.} \bibnamefont{Solnyshkov}},
  \bibinfo{author}{\bibfnamefont{P.}~\bibnamefont{Senellart}},
  \bibinfo{author}{\bibfnamefont{I.}~\bibnamefont{Sagnes}},
  \bibinfo{author}{\bibfnamefont{A.}~\bibnamefont{Lema\^itre}},
  \bibinfo{author}{\bibfnamefont{G.}~\bibnamefont{Malpuech}}, \bibnamefont{and}
  \bibinfo{author}{\bibfnamefont{J.}~\bibnamefont{Bloch}},
  \bibinfo{journal}{Phys. Rev. Lett.} \textbf{\bibinfo{volume}{106}},
  \bibinfo{pages}{126401} (\bibinfo{year}{2011}).

\bibitem[{\citenamefont{Christmann et~al.}(2012)\citenamefont{Christmann, Tosi,
  Berloff, Tsotsis, Eldridge, Hatzopoulos, Savvidis, and
  Baumberg}}]{christmann12a}
\bibinfo{author}{\bibfnamefont{G.}~\bibnamefont{Christmann}},
  \bibinfo{author}{\bibfnamefont{G.}~\bibnamefont{Tosi}},
  \bibinfo{author}{\bibfnamefont{N.~G.} \bibnamefont{Berloff}},
  \bibinfo{author}{\bibfnamefont{P.}~\bibnamefont{Tsotsis}},
  \bibinfo{author}{\bibfnamefont{P.~S.} \bibnamefont{Eldridge}},
  \bibinfo{author}{\bibfnamefont{Z.}~\bibnamefont{Hatzopoulos}},
  \bibinfo{author}{\bibfnamefont{P.~G.} \bibnamefont{Savvidis}},
  \bibnamefont{and} \bibinfo{author}{\bibfnamefont{J.~J.}
  \bibnamefont{Baumberg}}, \bibinfo{journal}{Phys. Rev. B}
  \textbf{\bibinfo{volume}{85}}, \bibinfo{pages}{235303}
  (\bibinfo{year}{2012}).

\bibitem[{\citenamefont{Dominici et~al.}(2013)\citenamefont{Dominici, Petrov,
  Matuszewski, Ballarini, Giorgi, Colas, Cancellieri, Fern\'andez, Bramati,
  Gigli et~al.}}]{arXiv_dominici13a}
\bibinfo{author}{\bibfnamefont{L.}~\bibnamefont{Dominici}},
  \bibinfo{author}{\bibfnamefont{M.}~\bibnamefont{Petrov}},
  \bibinfo{author}{\bibfnamefont{M.}~\bibnamefont{Matuszewski}},
  \bibinfo{author}{\bibfnamefont{D.}~\bibnamefont{Ballarini}},
  \bibinfo{author}{\bibfnamefont{M.~D.} \bibnamefont{Giorgi}},
  \bibinfo{author}{\bibfnamefont{D.}~\bibnamefont{Colas}},
  \bibinfo{author}{\bibfnamefont{E.}~\bibnamefont{Cancellieri}},
  \bibinfo{author}{\bibfnamefont{B.~S.} \bibnamefont{Fern\'andez}},
  \bibinfo{author}{\bibfnamefont{A.}~\bibnamefont{Bramati}},
  \bibinfo{author}{\bibfnamefont{G.}~\bibnamefont{Gigli}},
  \bibnamefont{et~al.}, \bibinfo{journal}{arXiv:1309.3083}
  (\bibinfo{year}{2013}).

\bibitem[{\citenamefont{Rossbach et~al.}(2013)\citenamefont{Rossbach, Levrat,
  Feltin, Carlin, Butt\'e, and Grandjean}}]{rossbach13a}
\bibinfo{author}{\bibfnamefont{G.}~\bibnamefont{Rossbach}},
  \bibinfo{author}{\bibfnamefont{J.}~\bibnamefont{Levrat}},
  \bibinfo{author}{\bibfnamefont{E.}~\bibnamefont{Feltin}},
  \bibinfo{author}{\bibfnamefont{J.-F.} \bibnamefont{Carlin}},
  \bibinfo{author}{\bibfnamefont{R.}~\bibnamefont{Butt\'e}}, \bibnamefont{and}
  \bibinfo{author}{\bibfnamefont{N.}~\bibnamefont{Grandjean}},
  \bibinfo{journal}{Phys. Rev. B} \textbf{\bibinfo{volume}{88}},
  \bibinfo{pages}{165312} (\bibinfo{year}{2013}).

\bibitem[{\citenamefont{Cilibrizzi et~al.}(2014)\citenamefont{Cilibrizzi,
  Ohadi, Ostatnicky, Askitopoulos, Langbein, and Lagoudakis}}]{cilibrizzi14a}
\bibinfo{author}{\bibfnamefont{P.}~\bibnamefont{Cilibrizzi}},
  \bibinfo{author}{\bibfnamefont{H.}~\bibnamefont{Ohadi}},
  \bibinfo{author}{\bibfnamefont{T.}~\bibnamefont{Ostatnicky}},
  \bibinfo{author}{\bibfnamefont{A.}~\bibnamefont{Askitopoulos}},
  \bibinfo{author}{\bibfnamefont{W.}~\bibnamefont{Langbein}}, \bibnamefont{and}
  \bibinfo{author}{\bibfnamefont{P.}~\bibnamefont{Lagoudakis}},
  \bibinfo{journal}{Phys. Rev. Lett.} \textbf{\bibinfo{volume}{113}},
  \bibinfo{pages}{103901} (\bibinfo{year}{2014}).

\bibitem[{\citenamefont{Utsunomiya et~al.}(2008)\citenamefont{Utsunomiya, Tian,
  Roumpos, Lai, Kumada, Fujisawa, Kuwata-Gonokami, L\"offler, H\"ofling,
  Forchel et~al.}}]{utsunomiya08a}
\bibinfo{author}{\bibfnamefont{S.}~\bibnamefont{Utsunomiya}},
  \bibinfo{author}{\bibfnamefont{L.}~\bibnamefont{Tian}},
  \bibinfo{author}{\bibfnamefont{G.}~\bibnamefont{Roumpos}},
  \bibinfo{author}{\bibfnamefont{C.~W.} \bibnamefont{Lai}},
  \bibinfo{author}{\bibfnamefont{N.}~\bibnamefont{Kumada}},
  \bibinfo{author}{\bibfnamefont{T.}~\bibnamefont{Fujisawa}},
  \bibinfo{author}{\bibfnamefont{M.}~\bibnamefont{Kuwata-Gonokami}},
  \bibinfo{author}{\bibfnamefont{A.}~\bibnamefont{L\"offler}},
  \bibinfo{author}{\bibfnamefont{S.}~\bibnamefont{H\"ofling}},
  \bibinfo{author}{\bibfnamefont{A.}~\bibnamefont{Forchel}},
  \bibnamefont{et~al.}, \bibinfo{journal}{Nat. Phys.}
  \textbf{\bibinfo{volume}{4}}, \bibinfo{pages}{700} (\bibinfo{year}{2008}).

\bibitem[{\citenamefont{Kohnle et~al.}(2011)\citenamefont{Kohnle, L\'eger,
  Wouters, Richard, Portella-Oberli, and Deveaud-Pl\'edran}}]{kohnle11a}
\bibinfo{author}{\bibfnamefont{V.}~\bibnamefont{Kohnle}},
  \bibinfo{author}{\bibfnamefont{Y.}~\bibnamefont{L\'eger}},
  \bibinfo{author}{\bibfnamefont{M.}~\bibnamefont{Wouters}},
  \bibinfo{author}{\bibfnamefont{M.}~\bibnamefont{Richard}},
  \bibinfo{author}{\bibfnamefont{M.~T.} \bibnamefont{Portella-Oberli}},
  \bibnamefont{and}
  \bibinfo{author}{\bibfnamefont{B.}~\bibnamefont{Deveaud-Pl\'edran}},
  \bibinfo{journal}{Phys. Rev. Lett.} \textbf{\bibinfo{volume}{106}},
  \bibinfo{pages}{255302} (\bibinfo{year}{2011}).

\bibitem[{\citenamefont{Almand-Hunter et~al.}(2014)\citenamefont{Almand-Hunter,
  Li, Cundiff, Mootz, Kira, and Koch}}]{almandhunter14a}
\bibinfo{author}{\bibfnamefont{A.~E.} \bibnamefont{Almand-Hunter}},
  \bibinfo{author}{\bibfnamefont{H.}~\bibnamefont{Li}},
  \bibinfo{author}{\bibfnamefont{S.~T.} \bibnamefont{Cundiff}},
  \bibinfo{author}{\bibfnamefont{M.}~\bibnamefont{Mootz}},
  \bibinfo{author}{\bibfnamefont{M.}~\bibnamefont{Kira}}, \bibnamefont{and}
  \bibinfo{author}{\bibfnamefont{S.~W.} \bibnamefont{Koch}},
  \bibinfo{journal}{Nature} \textbf{\bibinfo{volume}{506}},
  \bibinfo{pages}{471} (\bibinfo{year}{2014}).

\end{thebibliography}

\end{document}